\documentclass[a4paper,12pt]{article}
\usepackage{amsmath}
\usepackage{color} 
\usepackage{amssymb}
\usepackage{amsfonts}
\usepackage{graphicx}
\usepackage{epstopdf} 
\usepackage[english]{babel}
\usepackage[square]{natbib}

\title{Calibration and validation of a genetic regulatory network model describing the production of the gap gene protein Hunchback in \emph{Drosophila}  early development}         
\author{R. Dil\~ao and D. Muraro}        
\date{\today}          

\begin{document}

\maketitle

\begin{center}
Nonlinear Dynamics Group, Instituto Superior T\'ecnico,\\ Av. Rovisco Pais, 1049-001 Lisbon, Portugal.
\end{center} 

\begin{center}
rui@sd.ist.utl.pt, muraro@sd.ist.utl.pt 
\end{center}

\abstract{We fit the parameters of a differential equations model describing the production of gap gene proteins Hunchback and Knirps along the antero-posterior axis of the embryo of \emph{Drosophila}.
As initial data for the differential equations model, we take the antero-posterior distribution of
the proteins Bicoid,  Hunchback  and Tailless   at the beginning of  cleavage cycle 14. 
We calibrate and validate the model with experimental data using single- and multi-objective evolutionary optimization techniques.  In the multi-objective optimization technique, we  compute the associated Pareto fronts. 
We analyze the cross regulation mechanism between the gap-genes protein pair Hunchback-Knirps and
we show that the posterior distribution of Hunchback follow the experimental data if Hunchback is negatively regulated by the Huckebein protein.  This approach enables to predict the posterior localization on the embryo of the protein  Huckebein, and we validate with the experimental data the genetic regulatory network responsible for the antero-posterior distribution of the gap gene protein Hunchback. We discuss the importance of Pareto multi-objective optimization techniques in the calibration and validation of biological models.}

\vskip 0.5 cm
\noindent{\small{\sc KEYWORDS}: Genetic regulatory networks,  Hunchback-Knirps cross regulation, Huckebein.}
\vskip 1 cm

\section{Introduction}\label{intro}

In the \textit{Drosophila} egg, maternal mRNAs are placed near the poles of the oocyte by the mother's ovary cells, defining the antero-posterior axis of the embryo. Fertilization triggers the translation of these maternal mRNAs to proteins that regulate the expression of zygotic genes. Each of the zygotic genes is transcribed in certain regions of the embryo syncitial blastoderm, and the produced proteins act as transcription factors that regulate the expression of other zygotic genes. 

After fertilization, the first 13 nuclear divisions occur without the organization of cellular membranes, giving rise to a syncitial blastoderm. The cytoplasmic membranes only become completely formed three hours after fertilization, in the interphase following the $\hbox{14}^{th}$ mitotic cycle, just before the onset of gastrulation. 

During the syncitial stage, the transcribed zygotic genes are divided in three main families: gap, pair-rule and segment polarity genes. The proteins resulting from their expression define broad segmentation patterns along the antero-posterior axis of the embryo. These segmentation patterns appear as protein gradients along the  antero-posterior axis of the \textit{Drosophila} embryo,  \citep{F86, DN88, A87, N92}.

The proteins with origin in the maternal mRNAs  form gradients along the antero-posterior axis of the embryo. 
In the beginning of cleavage cycle 14,  proteins of maternal origin act as transcription factors for gap-genes, pair-rule and segment polarity genes.

There are several models aiming to describe proteins steady gradients in \emph{Drosophila} early development. 
Some models are based on the hypothesis of protein diffusion along the antero-posterior axis of the embryo, \citep{H05, AD06}, and other models are based on the diffusion of mRNA of maternal origin,
\citep{DM09, DMNS09}.
The protein diffusion hypothesis is sometimes justified by the absence of cellular membranes during the first 14 cleavage cycles of the embryo, and has been proposed by Nusslein-Volhard and co-workers in the late eighties,   \citep{DN88}. The mRNA diffusion hypothesis is supported by the recent observation of the mRNA Bicoid gradient, \citep{S09}, and the associated diffusion mechanism has been reported by
 \citep{Cha01} that observed rapid saltatory movements in injected mRNA \textit{bicoid}  with dispersion but without localization. 
 Another maternal mRNA (\textit{nanos}) has shown diffusive like behavior, \citep{FG03}.

Here, we will be concerned with the calibration and validation of the genetic regulatory network involving maternal proteins and the antero-posterior distribution of the gap genes Hunchback (HB) and Knirps (KNI) along the \textit{Drosophila} embryo. 
One of the reasons for this study is that  the regulation of the gradient of the HB protein in the posterior region of the embryo of \textit{Drosophila}  is poorly understood, \cite{M95}.

In order to calibrate the genetic regulatory network describing the production of the  gap genes  HB and KNI, we  make some remarks on the biological assumption of our approach.

\begin{description}
\item{1)} Models assuming that proteins of maternal origin diffuse along the embryo need the additional assumption that these proteins are continuously produced and degraded.
However this is unrealistic because: 
(i) Degradation has never been observed, \citep{KW07};
(ii) There are no proteins in the space around nuclei suggesting that protein do not diffuse, \citep{DM09};
(iii) Protein diffusion models need a condition on continuous production of mRNA of maternal origin,  
\citep{H05, AD06}, a feature that has never been observed. 
On the other hand,  models based on mRNA diffusion do not show these unrealistic features, and are able to produce accurately gradients of proteins of maternal origin, \citep{DM09, DMNS09}.
Anyway, the steady states obtained with the protein and the mRNA diffusion models  have the same functional form, (compare \cite{AD06} with \cite{DM09}), implying that the methodology followed here is not sensitive to the assumption
about the diffusion model for proteins or mRNA of maternal origin.

\item{2)} Threshold effects are important phenomena for the establishment of positional information in the embryo, \cite{W69}.
It has been shown that the mass action law models and the associated conservation laws lead naturally to stable gradients along the embryo of \textit{Drosophila}, without the need of \textit{ad-hoc} threshold or diffusive effects at the level of gap gene expression.  As a consequence, production models for gap gene proteins are simply described by ordinary differential equations models derived from the mass action law. Positional information is an emergent property  of the mass action associated conservation laws.
These biological assumptions have been tested qualitatively in \cite{AD05}, \cite{AD06} and \cite{DM09b}. 

\item{3)} The mechanism of protein production is described in two steps. 
In the first  step, we describe the establishment of the steady gradients of proteins produced from mRNAs with maternal origin.
In the second step, we consider that the maternal proteins are transcription factors for the gap-gene proteins. In order to simplify the model equations and the number of parameters for the description of the gap gene protein production, we assume that maternal origin proteins are not consumed in the activation or repression of the gap gene  proteins. In the case of the Hunchback protein, in a first step, we consider that the protein is produced from mRNA with  maternal origin. In a second step, it is assumed that HB is zygotically produced. In the initial gap gene phase, the gap gene proteins other than Hunchback are assumed to have zero initial concentration.  
\end{description}

This paper is organized as follows. In section \ref{mat}, we  fit the experimental data of maternal proteins Bicoid (BCD), Hunchback (HB) and Tailless (TLL) with the equations for the steady state of a reaction-diffusion based model. The biological assumptions made are the ones described above in 1). The experimental data were taken from the FlyEx database, \citep{Poustelnikova,Pisarev}, and the fits were obtained by an evolutionary search algorithm. 
In these fits, we reproduce accurately the experimental data for BCD, HB and TLL, and we determine along the antero-posterior axis of the embryo of \textit{Drosophila}  the initial localization of the mRNA of maternal origin.

In section \ref{gap}, we
introduce the graph of the genetic network associated with the production and the cross regulation of the gap gene proteins HB and Knirps (KNI) and we derive a mass action production model. Then, we describe the process of calibration of the parameters of the model with the experimental data.  The technique for parameter estimation is based on genetic algorithms with single- and multi-objective search  techniques.  As one of the main goals of this paper is to analyze the cross regulation of zygotically produced HB and KNI proteins, we have two  objectives to fulfill. In this context, we find a continuous set of parameter solutions or  Pareto front of the two-objectives optimization problem. This Pareto front corresponds to all possible admissible solutions of the bi-objective optimization problem. From the biological point of view,  all  the parameter  solutions on the Pareto front are admissible and they correspond to different instances of the model parameters. All these Pareto solutions are very close to the experimental data and this has been  evaluated by the chi-squared tests.

In section \ref{mm}, we describe the methodology of the multi-parameter fitting with evolutionary algorithms for one-objective and multi-objective optimization techniques. This section is essentially qualitative, describing the geometry and structure of the algorithms. All the computations are computationally involved and the programs are included in the supplementary material to this paper, \cite{SUP}.
Finally, in section \ref{conclusions}, we discuss the main conclusions of the paper.

\section{Results and Discussion}

\subsection{Steady state models for the distribution of proteins with maternal origin}\label{mat}

The first stages of the establishment of the positional information for the cellular differentiation of the \textit{Drosophila} embryo are determined by the initial distribution of maternal mRNAs and the  corresponding produced proteins. Here, we consider three proteins whose gradients are established prior to the gap gene phase. These three proteins are Bicoid (BCD), Hunchback (HB) and Tailless (TLL). 
We  fit the  steady state distribution of these proteins with the experimental data, taken from the FlyEx database 
\citep[http://flyex.ams.sunysb.edu/flyex/]{FlyEx1,FlyEx2,FlyEx3,Poustelnikova,Pisarev}. 
For the fits, we use a single-objective optimization technique for the distributions of BCD, HB and TLL.  
   
\textit{Hunchback} and \textit{bicoid} maternal mRNA are initially distributed along  the antero-posterior axis of the embryo. 
The \textit{tailless} gene is activated by the Torso (TOR) protein that has maternal origin. 
Here, we consider that TLL is produced directly from mRNA \textit{tll} which is not of maternal origin.
This choice is a simplification in the model  and the fit could be also obtained taking account of the activation of  the \textit{tll} gene by TOR, \citep{AD06}.

To describe the steady states of BCD, HB and TLL, we assume a model for the production of  proteins from the initial distribution of the associated mRNAs. In fact, we can adopt two alternative models. In one model, the produced protein diffuses and degrades along the embryo, leading to  a gradient like steady state, \citep{AD06}. In a second alternative model,  is the maternal mRNA that  diffuses and degrades, leading to a  gradient like steady state for the protein. The second model is experimentally supported by the fact the \textit{bicoid} mRNA shows a gradient, \citep{S09}.  It has been shown in \cite{DM09} that the protein steady states for both models have the same functional form, with parameters assuming different biological meanings. In the following, and without lack of generality, we assume the simple mRNA diffusion model for the production of proteins of maternal 
origin (Assumption 1) in \S~\ref{intro}).

In order to arrive at the steady state functional forms for the  distribution of BCD, HB and TLL proteins along the antero-posterior axis of  the \textit{Drosophila} embryo, we follow the mass action approach developed in \cite{AD05} and \cite{DM09b}. We consider the following kinetic diagrams for protein production,
\begin{eqnarray}
bcd & \stackrel{p_{BCD}}{\longrightarrow} & bcd + \hbox{BCD}\, , \qquad bcd \stackrel{d_{bcd}}{\longrightarrow} \nonumber\\
hb & \stackrel{p_{HB}}{\longrightarrow}
& hb + \hbox{HB}\, , \qquad  hb \stackrel{d_{hb}}{\longrightarrow} \nonumber \\
tll & \stackrel{p_{TLL}}{\longrightarrow} & tll + \hbox{TLL}\, , \qquad  tll \stackrel{d_{tll}}{\longrightarrow} \nonumber
\end{eqnarray}
where capital letters represent proteins and the italic letters the corresponding mRNAs. The constants $p_{BCD}$, $p_{HB}$ and $p_{TLL}$ are the protein production rates from mRNAs, and $d_{bcd}$,  $d_{hb}$ and $d_{tll}$  are  mRNA degradation rates. By the mass action law, to the above kinetic diagrams correspond the equations for the concentrations, 
\begin{eqnarray} 
\frac{\partial bcd}{\partial t} & = & -d_{bcd} bcd(x) +  D_{bcd}\frac{\partial^2 bcd}{\partial x^2}\label{eq1}\\ 
\displaystyle
\frac{\partial \hbox{BCD}}{\partial t} & = & p_{BCD} bcd(x)\label{eq2}\\  
\frac{\partial hb}{\partial t}   & = & -d_{hb} hb + D_{hb} \frac{\partial^2{hb}}{\partial{x^2}} \label{hb_eq}\label{eq3}\\
\frac{\partial \hbox{HB}}{\partial t}   & = & p_{HB} hb\label{HB_eq}\label{eq4}\\
\frac{\partial tll}{\partial t}   & = & -d_{tll} tll + D_{tll} \frac{\partial^2{tll}}{\partial{x^2}}\label{tll_eq}\label{eq5}\\
\frac{\partial \hbox{TLL}}{\partial t}   & = & p_{TLL} tll\label{eq6}
\end{eqnarray}
This system of differential equations describe the production and distribution of proteins and mRNA along the antero-posterior axis of the embryo of \textit{Drosophila}. The antero-posterior axis is described by the independent coordinate $x$. The $x$-dependent diffusion terms do not follow from the mass action law, but they have been added in order to describe the diffusive motion of the mRNAs.  The diffusion constants of the mRNAs are $D_{bcd}$, $D_{hb}$ and $D_{tll}$.

In order to solve this system of equations (\ref{eq1})-(\ref{eq6}), we now  define boundary and initial conditions. 
Denoting  by $L$ the length of the embryo, we have that $x\in [ 0,L]$.
Assuming zero flux boundary conditions for mRNAs and proteins, we have,
\begin{eqnarray}\displaystyle
\frac{\partial bcd}{\partial x}(x=0,t)=0\, , &\displaystyle  \frac{\partial bcd}{\partial x}(x=L,t) = 0\, ,\label{bc_bcd}\\ \displaystyle
\frac{\partial \hbox{BCD}}{\partial x}(x=0,t) =0\, , &\displaystyle   \frac{\partial \hbox{BCD}}{\partial x}(x=L,t) = 0\label{bc_BCD}\\ \displaystyle
\frac{\partial hb}{\partial x}(x=0,t)=0\, , &\displaystyle  \frac{\partial hb}{\partial x}(x=L,t) = 0\, ,\label{bc_hb}\\ \displaystyle
\frac{\partial \hbox{HB}}{\partial x}(x=0,t) =0\, , &\displaystyle   \frac{\partial \hbox{HB}}{\partial x}(x=L,t) = 0\label{bc_HB}\\ \displaystyle
\frac{\partial tll}{\partial x}(x=0,t)=0\, , &\displaystyle  \frac{\partial tll}{\partial x}(x=L,t) = 0\, ,\label{bc_tll}\\ \displaystyle
\frac{\partial \hbox{TLL}}{\partial x}(x=0,t) =0\, , &\displaystyle   \frac{\partial \hbox{TLL}}{\partial x}(x=L,t) = 0\label{bc_TLL}
\end{eqnarray}
for every $t \ge 0$. As initial conditions, we take the piecewise constant functions,
\begin{equation}
\begin{array}{ll}
&bcd(x,t=0)=\left\{ \begin{array}{ll}
B > 0, & \textrm{if}\ 0 < L_1 < x < L_2 < L\\
0, & \textrm{otherwise}\\
\end{array} \right.
\\[10pt]
&\hbox{BCD}(x,t=0)= 0\\
&hb(x,t=0)=\left\{ \begin{array}{ll}
H > 0, & \textrm{if}\ 0 < M_1 < x < M_2 < L\\
0, & \textrm{otherwise}\\
\end{array} \right.
\\[10pt]
&\hbox{HB}(x,t=0)= 0\\
&tll(x,t=0)=\left\{ \begin{array}{ll}
T_1 > 0, & \textrm{if}\ 0 < N_1 < x < N_2 < N_3\\
T_2 > 0, & \textrm{if}\  L_3 < x < L_N < L\\
0, & \textrm{otherwise}\\
\end{array} \right.
\\[10pt]
&\hbox{TLL}(x,t=0)= 0\\
\end{array}
\label{initialconditions}
\end{equation}
for every $x \in [ 0, L]$.  
The functions $bcd(x,t=0)$ and $hb(x,t=0)$ describe the distribution of \textit{bcd} and \textit{hb} maternal mRNA in the regions $[L_1, L_2 ]$ and $[M_1, M_2 ]$, respectively, of the antero-posterior axis of the embryo of \textit{Drosophila}. The function 
$tll(x,t=0)$ is the distribution of the hypothetical  \textit{tll} maternal mRNA in the region $[ N_1, N_2 ]  \cup    [ N_3, N_4 ]$, and $B$, $H$, $T_1$ and $T_2$ are constants.

Equations (\ref{eq1})-(\ref{eq6}), with boundary  conditions (\ref{bc_bcd})-(\ref{bc_TLL}), and initial conditions (\ref{initialconditions}) define the mRNA diffusion model for BCD, HB and TLL production. This model is linear, and the steady states $\hbox{BCD}_{eq}(x)$, $\hbox{HB}_{eq}(x)$ and $\hbox{TLL}_{eq}(x)$ can be obtained explicitly (for details see \cite{DM09}):
\begin{eqnarray}
\hbox{BCD}_{eq}(x) & = & 2\frac{a_1}{e^{2 a_2 / L} - 1}\cosh{\Big(a_2 \frac{x}{L}\Big)}
\Big( \sinh{\Big(a_2 \frac{L_2}{L}\Big)}-\sinh{\Big(a_2 \frac{L_1}{L}\Big)} \Big)\nonumber\\
& + & \frac{a_1}{2}\Big( e^{-a_2 (x+L_1)/L} - e^{-a_2 (x+L_2)/L} \Big) + I_{bcd}(x)\label{eq_bcd}\\
\hbox{HB}_{eq}(x) & = & 2\frac{a_3}{e^{2 a_4 / L} - 1}\cosh{\Big(a_4 \frac{x}{L}\Big)}
\Big( \sinh{\Big(a_4 \frac{M_2}{L}\Big)}-\sinh{\Big(a_4 \frac{M_1}{L}\Big)} \Big)\nonumber\\
& + & \frac{a_3}{2}\Big( e^{-a_4 (x+M_1)/L} - e^{-a_4 (x+M_2)/L} \Big) + I_{hb}(x)\label{eq_hb}\\
\hbox{TLL}_{eq}(x) & = & 2\frac{a_5}{e^{2 a_6 / L} - 1}\cosh{\Big(a_6 \frac{x}{L}\Big)}
\Big( \sinh{\Big(a_6 \frac{N_2}{L}\Big)}-\sinh{\Big(a_6 \frac{N_1}{L}\Big)} \Big)\nonumber\\
& + & \frac{a_5}{2}\Big( e^{-a_6 (x+N_1)/L} - e^{-a_6 (x+N_2)/L} \Big) + I_{1tll}(x)\nonumber\\
& + & 2\frac{a_7}{e^{2 a_8 / L} - 1}\cosh{\Big(a_8 \frac{x}{L}\Big)}
\Big( \sinh{\Big(a_8 \frac{N_4}{L}\Big)}-\sinh{\Big(a_8 \frac{N_3}{L}\Big)} \Big)\nonumber\\
& + & \frac{a_7}{2}\Big( e^{-a_8 (x+N_3)/L} - e^{-a_8 (x+N_4)/L} \Big) + I_{2tll}(x)
\label{eq_tll} 
\end{eqnarray} 
where,
\begin{eqnarray}\displaystyle
I_{bcd}(x)&=&\left\{ \begin{array}{ll}
\frac{a_1}{2}\Big( e^{-a_2 (L_1 - x)/L} - e^{-a_2 (L_2 - x)/L} \Big), & \textrm{if}\ x < L_1\\
a_1 - \frac{a_1}{2}\Big( e^{-a_2 (x - L_1)/L} + e^{-a_2 (L_2 - x)/L} \Big), & \textrm{if}\ L_1 \le x \le L_2\\
\frac{a_1}{2}\Big( e^{-a_2 (x - L_2)/L} - e^{-a_2 (x - L_1)/L} \Big), & \textrm{if}\ x > L_2
\end{array} \right. \nonumber \\
 \label{Ibcd} \\
I_{hb}(x)&=&\left\{ \begin{array}{ll}
\frac{a_3}{2}\Big( e^{-a_4 (M_1 - x)/L} - e^{-a_4 (M_2 - x)/L} \Big), & \textrm{if}\ x <M_1\\
a_3 - \frac{a_3}{2}\Big( e^{-a_4 (x - M_1)/L} + e^{-a_4 (M_2 - x)/L} \Big), & \textrm{if}\ M_1 \le x \le M_2\\
\frac{a_3}{2}\Big( e^{-a_4 (x - M_2)/L} - e^{-a_4 (x - M_1)/L} \Big), & \textrm{if}\ x > M_2
\end{array} \right. \nonumber \\
 \label{Ihb} \\
I_{1tll}(x)&=& \left\{ \begin{array}{ll}
\frac{a_5}{2}\Big( e^{-a_6 (N_1 - x)/L} - e^{-a_6 (N_2 - x)/L} \Big), & \textrm{if}\ x < N_1\\
a_5 - \frac{a_5}{2}\Big( e^{-a_6 (x - N_1)/L} + e^{-a_6 (N_2 - x)/L} \Big), & \textrm{if}\ N_1 \le x \le N_2\\
\frac{a_5}{2}\Big( e^{-a_6 (x - N_2)/L} - e^{-a_6 (x - N_1)/L} \Big), & \textrm{if}\ x > N_2
\end{array} \right. \nonumber \\
\label{I1tll} \\
I_{2tll}(x)&=& \left\{ \begin{array}{ll}
\frac{a_7}{2}\Big( e^{-a_8 (N_3 - x)/L} - e^{-a_8 (N_4 - x)/L} \Big), & \textrm{if}\ x < N_3\\
a_7 - \frac{a_7}{2}\Big( e^{-a_8 (x - N_3)/L} + e^{-a_8 (N_4 - x)/L} \Big), & \textrm{if}\ N_3 \le x \le N_4\\
\frac{a_7}{2}\Big( e^{-a_8 (x - N_4)/L} - e^{-a_8 (x - N_3)/L} \Big), & \textrm{if}\ x > N_4
\end{array}\right.  \nonumber \\
 \label{I2tll} 
\end{eqnarray}
and, 
\begin{eqnarray}
a_1 &=& B\frac{p_{BCD}}{d_{bcd}},\qquad a_2^2 = d_{bcd} \frac{L^2}{D_{bcd}}\label{a1}\\
a_3 &=& H\frac{p_{HB}}{d_{hb}},\qquad a_4^2 = d_{hb} \frac{L^2}{D_{hb}}\label{a2}\\
a_5 &=& T_1\frac{p_{TLL}}{d_{tll}},\qquad a_6^2 = d_{tll} \frac{L^2}{D_{tll}}\label{a3}\\
a_7 &=& T_2\frac{p_{TLL}}{d_{tll}},\qquad a_8^2 = d_{tll} \frac{L^2}{D_{tll}}\label{a4}
\end{eqnarray}
Note that $a_6=a_8$.

\begin{figure}[htpb]
\centering
\includegraphics[width=9cm]{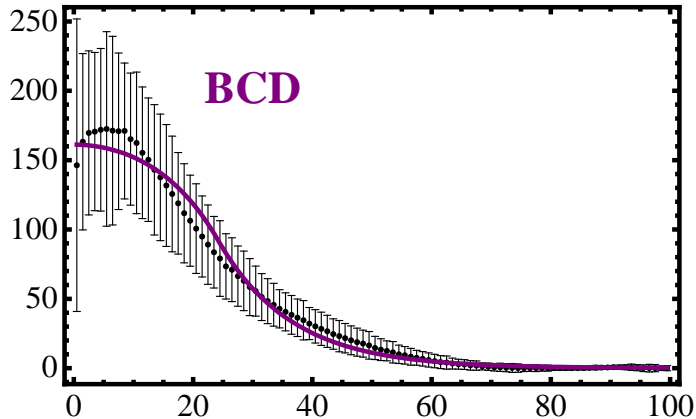}
\caption{Dots and error bars represent the mean values and the standard deviations  of the concentration of the protein Bicoid (BCD)  along the antero-posterior axis of the embryo of \textit{Drosophila}, at  cleavage cycle 14A.  
The fit has been obtained with the steady state solution defined in (\ref{eq_bcd}),  (\ref{Ibcd}) and (\ref{a1}). The parameter values found in the fit are:
$L_1 = 0.00$,  $L_2 = 0.24$,    $a_{1} = 186.83$ and $ a_{2} = 8.18$. The reduced chi-squared value of this fit is $\chi_{BCD}^2(\vec{p_1}) =0.13$. The interval $[L_1,L_2]$ is the region where mRNA \textit{bcd} is deposited by the mother's ovary cells.}
\label{fig1}
\end{figure}

\begin{figure}[htpb]
\centering
\includegraphics[width=9cm]{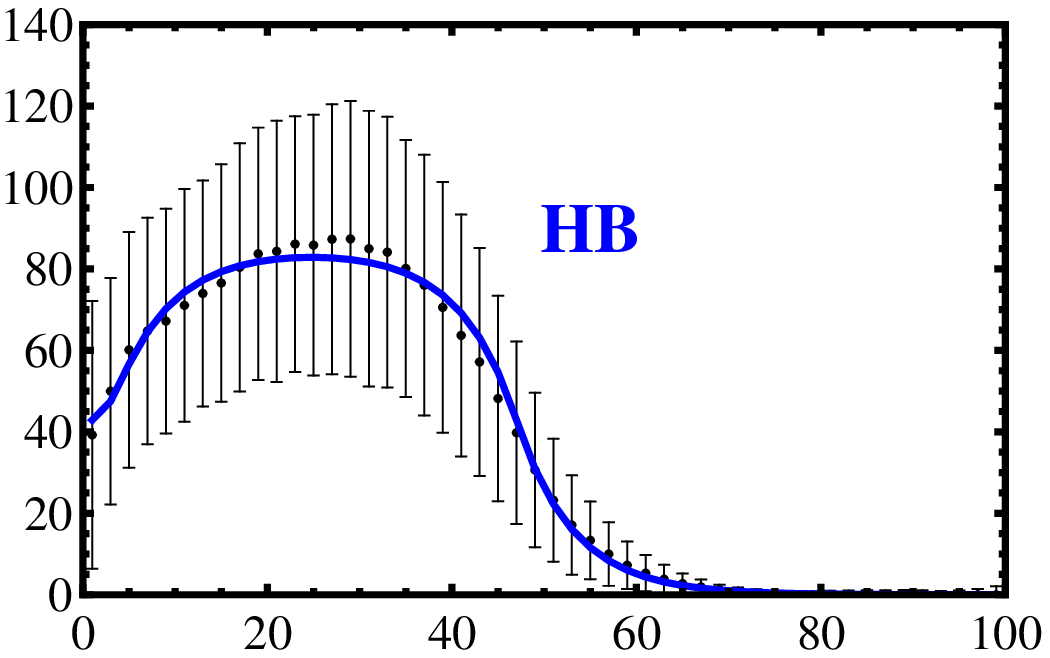}
\caption{Dots and error bars represent the mean values and the standard deviations  of the concentration of the protein Hunchback (HB)  along the antero-posterior axis of the embryo of \textit{Drosophila}, at the end of cleavage cycle 13. 
The fit has been obtained with the steady state solution defined in (\ref{eq_hb}),  (\ref{Ihb}) and (\ref{a2}).   The parameter values found in the fit are:
$M_{1} = 0.04$, $ M_{2} = 0.47$, $a_{3} = 85.09$ and $ a_{4} = 16.36$. The reduced chi-squared value of this fit is $\chi_{HB}^2(\vec{p_2}) =0.02$. The interval $[M_1,M_2]$ is the region where mRNA \textit{hb} is deposited by the mother's ovary cells.}
\label{fig2}
\end{figure}

\begin{figure}[htpb]
\centering
\includegraphics[width=9cm]{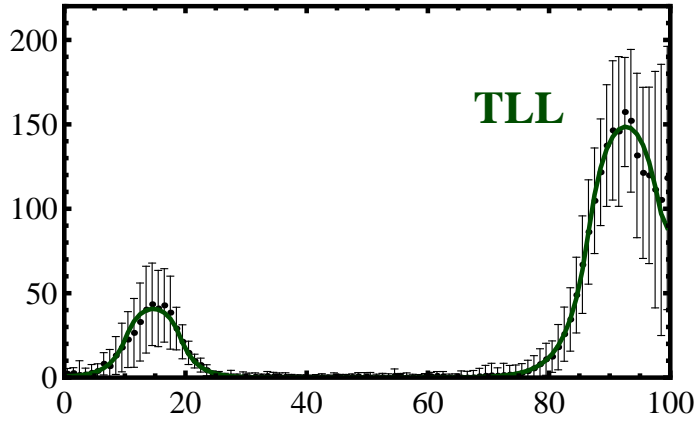}
\caption{Dots and error bars represent the mean values and the standard deviations  of the concentration of the protein Tailless (TLL)  along the antero-posterior axis of the embryo of \textit{Drosophila}, at  cleavage cycle 14A.  
The fit has been obtained with the steady state solution defined in (\ref{eq_tll}),  (\ref{I1tll}),   (\ref{I2tll}), (\ref{a3})  and (\ref{a4}).    The parameter values found in the fit are:
$N_{1} = 0.10$, $N_{2} = 0.19$, $ N_{3} = 0.86$, $ N_{4} = 0.97$, 
$a_{5} = 49.05$, $ a_{6} =39.56$,  $ a_{7} = 175.83$ and $ a_{8} = 30.33$. The reduced chi-squared value of this fit is $\chi_{TLL}^2(\vec{p_3}) =0.03$.}
\label{fig3}
\end{figure}

The steady states for the gradients of proteins BCD, HB and TLL are given by equations (\ref{eq_bcd})-(\ref{a4}). 
For the calibration of equations (\ref{eq_bcd})-(\ref{a4})  with the experimental data, we have taken from the FlyEx database the mean  antero-posterior distributions of the proteins BCD, HB and TLL. 
These distributions have been calculated from the individual spatial distributions measured in $954$ different embryos.
These distributions are assumed to  correspond to a steady state and, in the case of HB, the steady state is assumed to be established at the  end of cleavage cycle 13. For the BCD and the TLL proteins, the steady state  distribution corresponds to the beginning of cleavage cycle 14A.  
 In Figs. \ref{fig1},  \ref{fig2}   and   \ref{fig3},  we show the mean values and the corresponding standard deviations of the gradients of proteins BCD, HB and TLL along the antero-posterior axis of the embryo of \textit{Drosophila}. In these pictures, all the embryos have been scaled to the length $L=100$. 

To fit the experimental data of  BCD, HB and TLL with (\ref{eq_bcd})-(\ref{a4}), we have used an  evolutionary search algorithm (see \S \ref{CMA-ES}), and the choice of parameters has been done by minimizing the 
 reduced chi-square functions, 
\begin{eqnarray}
\begin{array}{rcl}\displaystyle
\chi_{BCD}^2(\vec{p_1}) &=& \displaystyle \frac{1}{n} \sum_{i=1}^{n}\frac{(\hbox{BCD}_{eq}(x_i,\vec{p_1}) - \hbox{BCD}_{mean}(x_i))^2}{\hbox{BCD}_{\sigma^2}(x_i)}\\ \displaystyle
\chi_{HB}^2(\vec{p_2}) &=& \displaystyle \frac{1}{n} \sum_{i=1}^{n}\frac{(\hbox{HB}_{eq}(x_i,\vec{p_2}) - \hbox{HB}_{mean}(x_i))^2}{\hbox{HB}_{\sigma^2}(x_i)} \\ \displaystyle
\chi_{TLL}^2(\vec{p_3}) &=& \displaystyle \frac{1}{n} \sum_{i=1}^{n}\frac{(\hbox{TLL}_{eq}(x_i,\vec{p_3}) - \hbox{TLL}_{mean}(x_i))^2}{\hbox{TLL}_{\sigma^2}(x_i)}
\end{array}
\label{chi}
\end{eqnarray}
where $\vec{p_1}=(L_1,L_2, a_1, a_2)$ is the vector of the free parameters for the BCD production model, 
$\vec{p_2}=(M_1,M_2, a_3, a_4)$ is the vector of the free parameters for the HB production model,
and 
\[
\vec{p_3}=(N_1,N_2, N_3, N_4, a_5, a_6, a_7, a_8)
\]
is the vector of the free parameters for the TLL production model. The functions $\hbox{BCD}_{mean}(x)$, $\hbox{HB}_{mean}(x)$ and $\hbox{TLL}_{mean}(x)$ are the mean values of the protein concentrations along the antero-posterior axis of the embryo, and the functions ${\hbox{BCD}_{\sigma^2}(x)}$, ${\hbox{HB}_{\sigma^2}(x)}$ and ${\hbox{TLL}_{\sigma^2}(x)}$ are
the associated standard deviations. 
 In the fits, we have assumed  that $a_6$ and $a_8$ are independent parameters and we have taken $n=100$. This assumption gives more plasticity to the data fitting and is based on the assumption that the goal of the fits is to find an accurate fitting function for TLL. The protein TLL is activated by the maternal origin protein Torso and this mechanism is not considered here, \citep{AD06}. 
The results of the three calibrations are shown in Figs.~\ref{fig1},   \ref{fig2}  and   \ref{fig3}, and the fitted parameter values are listed in the figure captions. 

From the fits in Figs.~\ref{fig1},   \ref{fig2}  and   \ref{fig3}, we conclude that the steady state model describes well the distribution of proteins predicted from the mRNAs with maternal origin. The values of the reduced chi-squared test show that the agreement between data and fits are very good. If a model is successfully calibrated with experimental data, then it corresponds, with some degree of plausibility, to the mechanism that it pretends to describe.

As already stated, the steady state solutions (\ref{eq_bcd})-(\ref{eq_tll}) are functionally similar to the ones obtained with the protein reaction-diffusion model, compare equation (10) of \cite{AD06} with equation (4) of \cite{DM09} . 

Programs and software tools for evolutionary algorithms optimization techniques and model construction and analysis are available in the supplementary material, \cite{SUP}.

We are now in condition to make the calibration and validation of the gap gene proteins HB and KNI.

\subsection{Fitting the gap-genes}\label{gap}

To describe the production of gap gene proteins,  we consider that BCD, HB and TLL proteins are in the steady state with a gradient like distribution along the antero-posterior axis of the embryo of \textit{Drosophila}, Figs.~\ref{fig1},   \ref{fig2}  and   \ref{fig3}.
We consider that the production of the gap-genes proteins begins at the cleavage cycle 14 and, at this stage, we do not consider diffusion (Assumption 2) in \S~\ref{intro}).
We expect that the positional information is obtained by a threshold mechanism with diffusion playing no role, \citep{AD05, DM09b}.
So, to model the gap-gene transcriptional regulation of Hunchback (HB) and Knirps (KNI), we take as initial conditions the antero-posterior distribution of BCD, HB and TLL, as found in the previous section. 
Then, we build the regulatory network following the mass action law strategy of \cite{AD05} and \cite{DM09b}.

The basic pattern of gap-genes HB and KNI expression pattern is due to strong mutual repression between these genes.
This complementarity is particularly clear in the experimental data for the couple HB-KNI  at cleavage cycle 14A-4, and has been confirmed in \cite{JR06} and earlier results, together with the repression of TLL over KNI, affecting the posterior pole of the embryo.    

\begin{figure}[htpb]
\centering
\includegraphics[width=4cm]{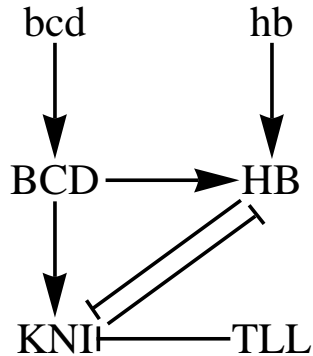}
\caption{Genetic regulatory network graph associated with the cross regulation of the proteins Hunchback (HB) and Knirps (KNI) in \textit{Drosophila} early development. The protein KNI is activated in the embryo by Bicoid (BCD). HB has a maternal origin and is also regulated by BCD. Both KNI and zygotically produced HB repress each other. In Fig.~\ref{fig2}, we show the distribution of HB at the end of the maternal phase, before considering the regulation by BCD as described in this genetic network graph.}
\label{fig4}
\end{figure}

The gap-gene genetic regulatory network involving HB and KNI is displayed in Fig.~\ref{fig4}.  Associated with the regulatory network of Fig.~\ref{fig4}, we  build the model for this genetic regulatory model based on the mass action law and following the description of transcriptional regulation by the operon model and developed in \cite{AD05} and \cite{DM09b}. Using the \textit{Mathematica} software package \textit{GeneticNetworks.m}, we obtain the equations describing the time evolution of the gap gene protein concentrations. These differential equations involve the concentration of the proteins and of the gap genes with the different biding sites occupied or not. In the particular  case of  Fig.~\ref{fig4}, the full system of ordinary differential equations has 14 equations and 23 free parameters (\cite{SUP}).

In order to test the validity and completeness of the genetic regulatory network in Fig.~\ref{fig4},  we took from the FlyEx database the experimental data of the distribution of HB and KNI for the late cleavage cycle 14, and we
have integrated numerically  in \textit{Mathematica} the model equations generated by the  \textit{GeneticNetworks.m} software package.  
The free parameters on the model equations were determined with a bi-objective optimization technique (\S \ref{MO-CMA-ES}), minimizing the mean squared deviations between the model solutions and the experimental data. Denoting by $\hbox{HB}(x,t)$ and  $\hbox{KNI}(x,t)$ the solutions of the model equations, we have fitted the experimental data for the antero-posterior distribution of HB and KNI
with the functions  $\alpha_{hb}\hbox{HB}(x,t)$ and  $\alpha_{kni}\hbox{KNI}(x,t)$, where $\alpha_{hb}$ and $\alpha_{kni}$ are proportionality constants. The introduction of the proportionality constants $\alpha_{hb}$ and $\alpha_{kni}$ is due to the fact that experiments do not correspond to a direct measurement of local protein concentration, but it is proportional to protein concentration. These proportionality constants change from one protein to  another. With these two additional proportionality constants and time as a free parameter, we have fitted the $23$ parameters of the model with a bi-objective optimization technique and we have calculated the associated Pareto front.

\begin{figure}[htpb]
\centering
\includegraphics[width=12cm]{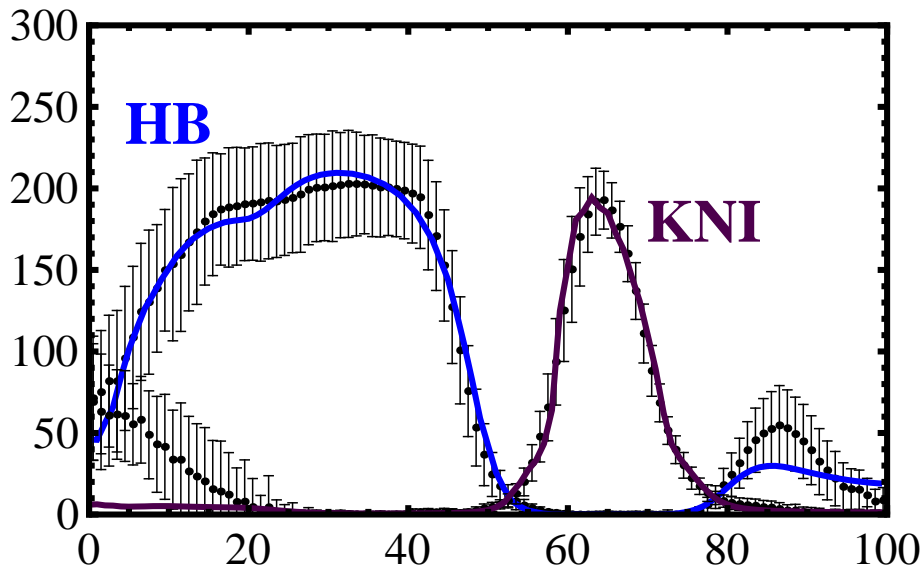}
\caption{Dots and error bars represent the mean values and the standard deviations  of the concentration of the proteins Hunchback (HB) and Knirps (KNI)  along the antero-posterior axis of the embryo of \textit{Drosophila}, at  the end of cleavage cycle 14A.  
The fit has been obtained by a multi-objective optimization technique as described in \S~\ref{MO-CMA-ES}. 
The continuous lines correspond to  the differential equation model solutions $\alpha_{hb}\hbox{HB}(x,t^*)$ and  $\alpha_{kni}\hbox{KNI}(x,t^*)$, away from the steady state ($t^*< \infty$), and for a particular set of parameter values localized on the Pareto front of the bi-objective optimized solution. 
In this case, the fitted value of time is $t^*=10$~s, and the fitted proportionality constants have the values $\alpha_{hb}=0.1$ and $\alpha_{kni}=2.0$. The penalized chi-squared values, (\ref{chi2}),  of these fits are $\chi_{HB}^2(\vec{p_4}) =0.28$ and $\chi_{KNI}^2(\vec{p_4}) =0.50$, where $p_4$ is the vector of the  parameters that have been fitted. In this case, $P=23$.
In this model, this shows clearly that the genetic regulatory network of Fig.~\ref{fig4} describes well the HB and KNI distribution away from the posterior tip of the  embryo of  \textit{Drosophila}.
}
\label{fig5}
\end{figure}

In Fig.~\ref{fig5}, we show this data and the corresponding fits.  From the fitts, it is shown clearly that the genetic regulatory network of Fig.~\ref{fig4} describes well the HB and the KNI  distributions away from the posterior tip of the  \textit{Drosophila} embryo. Clearly, complementarity of the proteins HB and KNI in the middle region of the embryo is observed.
This fact suggests that there are other genes that regulate the posterior region of the embryo. A plausible candidate is the Huckebein (HKB) protein, \cite{M95}.

\begin{figure}[htpb]
\centering
\includegraphics[width=4cm]{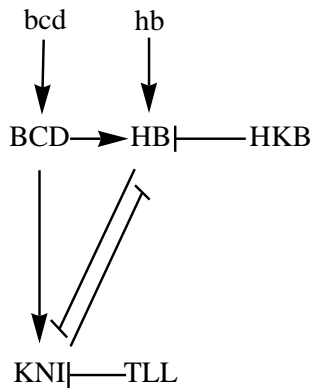}
\caption{Genetic regulatory network graph associated with the cross regulation of the proteins Hunchback (HB), Knirps (KNI) and Huckebein (HKB) in \textit{Drosophila} early development. The transcription repression of HKB on the transcription of HB is described in  \cite{M95}.}
\label{fig6}
\end{figure}

In order to analyze the distribution of HB near the posterior region of the embryo, there is  experimental evidence that Huckebein (HKB) protein has a band near the posterior pole of the embryo,  repressing the zygotic production of HB. Therefore, we introduce  HKB in the gap gene regulatory network as in Fig.~\ref{fig6}. As there is experimental evidence that HKB represses the production of HB near the posterior pole of the embryo, we assume a band type localization of HKB near the posterior tip of the embryo.

\begin{figure}[htpb]
\centering
\includegraphics[width=12cm]{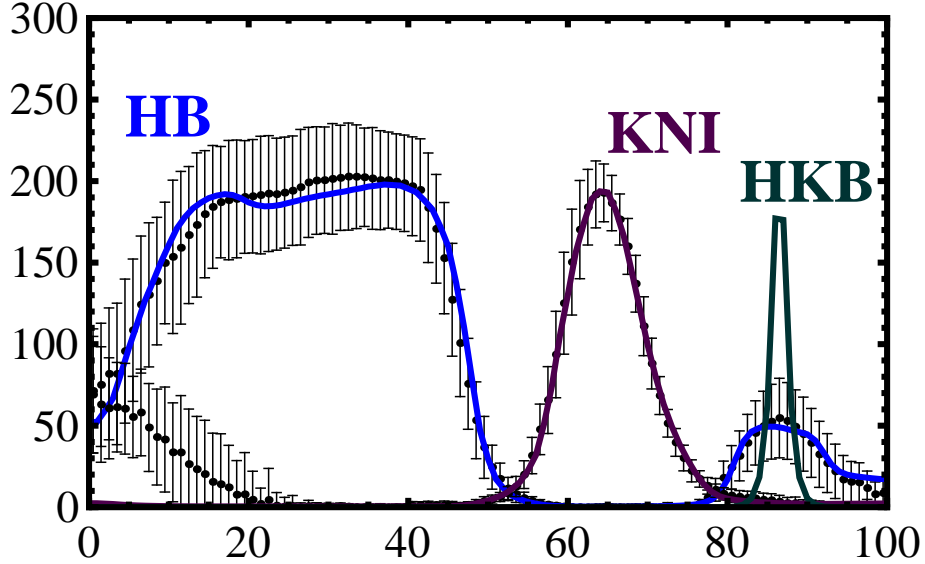}
\caption{Dots and error bars represent the mean values and the standard deviations  of the concentration of the proteins Hunchback (HB) and Knirps (KNI)  along the antero-posterior axis of the embryo of \textit{Drosophila}, at  the end of cleavage cycle 14A. Due to the lack of experimental data on HKB its spatial experimental distribution is not represented.   
The fit has been obtained by a bi-objective optimization technique for HB and KNI, having also as free the parameters  that describe the 
HKB distribution (\ref{hkb}).
The continuous lines correspond to  the differential equation model solutions $\alpha_{hb}\hbox{HB}(x,t^*)$, $\alpha_{kni}\hbox{KNI}(x,t^*)$ and $\hbox{HKB}_{eq}(x)$,  for a particular set of parameter values localized on the Pareto front of the bi-objective optimized solution. 
In the case, the fitted value of time is $t^*=29.1$~s, and the fitted proportionality constants have the values $\alpha_{hb}=0.11$ and $\alpha_{kni}=0.65$. 
The penalized chi-squared value, (\ref{chi2}), of this fit are $\chi_{HB}^2(\vec{p_5}) =0.14$ and $\chi_{KNI}^2(\vec{p_5}) =0.59$, where $p_5$ is the vector of the  parameters that have been fitted. In this case, $P=31$. The parameter values found for the prediction of the HKB distribution (\ref{hkb}) are:
$P_{1} = 0.856$, $ P_{2} = 0.873$, $ b_{1} = 296.74$ and $b_{2} = 121.87$.
The HKB distribution found in the fit shows the existence of a stripe of the protein HKB near the posterior pole of the embryo as suggested experimentaly. With this fit, it is clearly shown that the genetic regulatory network of Fig.~\ref{fig6} describes well the HB and KNI distribution along all the antero-posterior axis  of the  embryo of \textit{Drosophila}.}
\label{fig7}
\end{figure}

By  consistence with the model construction done in the previous section \S~\ref{mat}, we assume that the HKB protein is localized with the following steady state distribution,
\begin{eqnarray}
\hbox{HKB}_{eq}(x) & = & 2\frac{b_1}{e^{2 b_2 / L} - 1}\cosh{\Big(b_2 \frac{x}{L}\Big)}
\Big( \sinh{\Big(b_2 \frac{P_2}{L}\Big)}-\sinh{\Big(b_2 \frac{P_1}{L}\Big)} \Big)\nonumber\\
& + & \frac{b_1}{2}\Big( e^{-b_2 (x+P_1)/L} - e^{-b_2 (x+P_2)/L} \Big) + I_{bcd}(x)\label{hkb}
\end{eqnarray}
where $P_1$, $P_2$, $b_1$ and $b_2$ are constants to be fitted and have the same meaning as the  constants in the BCD equilibrium distribution  (\ref{eq_bcd}). Under these conditions, we have introduced the HKB distribution into the previous model and we have repeated the bi-objective optimization analysis and we have calculated the associated Pareto front. In  Fig.~\ref{fig7}, we show one of the Pareto instances of the fit of the experimental data,
as well as the fitted distribution of the protein HKB.

The quality of the fits in Figs.~\ref{fig5} and \ref{fig7} were evaluated from the penalized  chi-square functions, 
\begin{eqnarray}
\begin{array}{rcl}\displaystyle
\chi_{HB}^2(\vec{p}) &=& \displaystyle \frac{1}{n-P/2} \sum_{i=1}^{n}\frac{(\hbox{HB}_{eq}(x_i,\vec{p}) - \hbox{HB}_{mean}(x_i))^2}{\hbox{HB}_{\sigma^2}(x_i)}\\\displaystyle
\chi_{KNI}^2(\vec{p}) &=& \displaystyle \frac{1}{n-P/2} \sum_{i=1}^{n}\frac{(\hbox{KNI}_{eq}(x_i,\vec{p}) - \hbox{KNI}_{mean}(x_i))^2}{\hbox{KNI}_{\sigma^2}(x_i)} 
\end{array}
\label{chi2}
\end{eqnarray}
where $\vec{p}$ is the vector of the free parameters for the differential equation model and $P$ is the dimension of the vector $\vec{p}$.

From the fits in Fig.~\ref{fig7}, we conclude that the transcriptional cross repression of HB over KNI and the transcriptional repression of  HKB over HB describe well the spatial distributions of HB and KNI proteins along the antero-posterior axis  of the  embryo of \textit{Drosophila}. This result also predicts the distribution of the protein HKB. 

Another important conclusion common to both fits is that gap gene protein expression is a dynamic process with a very fast expression time, of the order of $30$~s (Fig.~\ref{fig7}). This expression time is calculated relative to the beginning of cleave stage 14A.

Programs and software tools for multi-objective optimization techniques and Pareto front solutions are available in the supplementary material, \citep{SUP}.

\section{Materials and Methods}\label{mm}

In this section, we briefly describe the algorithms that we have applied to calculate the parameters that best fit the experimental data to the model equations generated by the \textit{Mathematica} software package \textit{GeneticNetworks.m}. These algorithms are based on the {\em Covariance Matrix Adaptation Evolution Strategy} (CMA-ES) approach, an evolutionary algorithm for black-box continuous optimization, \citep{hansen01, hansen08}. The first algorithm is for single-objective optimization,  used in \S \ref{mat}, and will be referred by CMA-ES. The second algorithm is the multi-objective version of CMA-ES,  used in \S \ref{gap}, and uses several CMA-ES processes together with a global Pareto-dominance based selection, \citep{IHR07}. In a maximization or a minimization problem, there is a fitness function relative to which an optimization is found. In multi-objective
optimization problems, there are several fitness functions, and in general when we optimize in order to a fitness function, we are worsening in order to the other fitness function. Pareto optimization is  a way of obtaining 
optimal solutions that are not dominated in a certain sense by other solutions.

\subsection{Single-objective optimization: CMA-ES}\label{CMA-ES}

CMA-ES is  an evolutionary algorithm that uses a population of $\mu$ parents to generate $\lambda$ offspring, and deterministically selects the best $\mu$ of those $\lambda$ offspring for the next generation. 

To have an idea of the parameter identification search problem, we take first a compact subset $X$ of the parameter space $S$. The number of the parameter to be identified is the dimension of $S$. Set an initial point $p_0\in X\subset S$ and let $C=I_n$ be a covariance matrix, where $I_n$ is the $n\times n$ identity matrix. 
Then, from the  multivariate Gaussian distribution with covariance matrix $C$ and mean value $p_0$, sample  $\lambda$ offsprings. For each offspring calculate a fitness function, in our case the, chi-squared distributions (\ref{chi}). From the best $\mu$ ($<\lambda$) offsprings, according to the fitness function, recalculate a new mean value
$p_0$ and a new (unbiased estimator) covariance matrix $C$, and repeat the procedure. After several iterations, the best individual ever found is a candidate for the best choice of parameters. For details see \cite{hansen01} and  \cite{hansen08}. Maternal protein distributions in Figs.~\ref{fig1}, \ref{fig2} and \ref{fig3}
have been determined according to this technique.

\subsection{Pareto optimization}\label{Pareto}

Pareto optimization is concerned with the finding of the set of {\em optimal trade-offs} between conflicting objectives. Namely, Pareto solutions of a multi-objective problem are optimized solutions such that the value of one objective cannot be improved without degrading the value of at least another objective. Such best compromises are what is called the {\em Pareto set} of the multi-objective optimization problem.
Pareto optimization is based on the notion of \emph{dominance}.  
Consider a minimization problem with $M$ real valued objective functions $f=(f_1 ,\ldots,f_M)$ defined on a subset  $X\subset \mathbb{R}^n$. 
A solution of the optimization problem $\bar{x}\in X$ is said to \emph{dominate} another solution $x\in X$, denoted by $\bar{x}\prec x$, if,
\[
\begin{array}{rcl}
 \forall m & \in & \{ 1,\ldots,M \}:   \\
 (f_m (\bar{x}) &\le &f_m (x)  ) \land \left(\exists m\in \{ 1,\ldots,M \}: f_m (\bar{x})<f_m (x)\right) \, .
 \end{array}
\]
The Pareto set of an optimization problem is the set of nondominated solutions of a minimization (maximization) problem. More formally,
\[
 \hbox{Pareto Set}=\{ x: (x\in X) \land \not\exists \bar{x}\in X:  \bar{x}\prec x\}\, .
 \]
 The Pareto front is the image of the Pareto set in the fitness space,
\[
   \hbox{Pareto front}=\{ f(x): (x\in X) \land \not\exists \bar{x}\in X:  \bar{x}  \prec x\}\, .
\]   

The goal of Pareto optimization is to find the \emph{Pareto set} of optimized parameters 
and the Pareto front. Therefore, in a multi-objective approach, the natural choice for unbiased parameter estimation is the determination of the Pareto set of a given optimization problem. In this set, all the solutions are optimized solutions. The distributions  of the gap gene proteins HB and KNI in Figs.~\ref{fig5} and \ref{fig7} correspond to parameter values on a Pareto set of the bi-objective optimization problem. In general, all the solutions on the Pareto set are equally acceptable, \citep{DMNS09}.


\subsection{Multi-objective optimization: MO-CMA-ES}\label{MO-CMA-ES}

The Multi-Objective CMA-ES (MO-CMA-ES)  optimization technique is based on the specific CMA-ES algorithm with a random choice of a large number of  initial points in the search parameter space, \citep{IHR07} .  
Once defined the multidimensional parameter search space, $X$, we proceed with the multi-objective optimization technique to determine the Pareto set and Pareto front of the two fitting problems of \S~\ref{gap}.  The  MO-CMA-ES techniques can be divided in three steps:
\begin{description}
\item{1)}  In the compact search space $X$, choose randomly $\mu$ parents. For each parent, one offspring is generated with the CMA-ES algorithm. Initially, the CMA-ES algorithm is implemented with the identity as covariance matrix.
\item{2)}  We now rank the best $\mu$ individuals from the set of $2\mu$ individuals found previously. For that 
we use the concept of Pareto  dominance. From the   $2\mu$ individuals, we select the set of all the non-dominated individuals and we give them rank 1. We apply the same procedure to the remaining individuals and we obtain the rank 2 individuals, \citep{D02}. This procedure continues until a last rank is reached. 
\item{3)}   In order to rank the  individuals within the same rank of non-dominance found previously,  we do a second ranking of  individuals within each rank. This second order ranking is done  according to an \emph{hypervolume measure} in the objective space,  \citep{K03}. After this new ranking, we retain only the best $\mu$ individuals. With this procedure, we obtain an approximation to the Pareto front with an approximately uniform distribution of individuals within each rank. Then, we repeat these three procedures until a good converge to the Pareto front is achieved.
\end{description}

\begin{figure}[htpb]
\centering
\includegraphics[width=10cm]{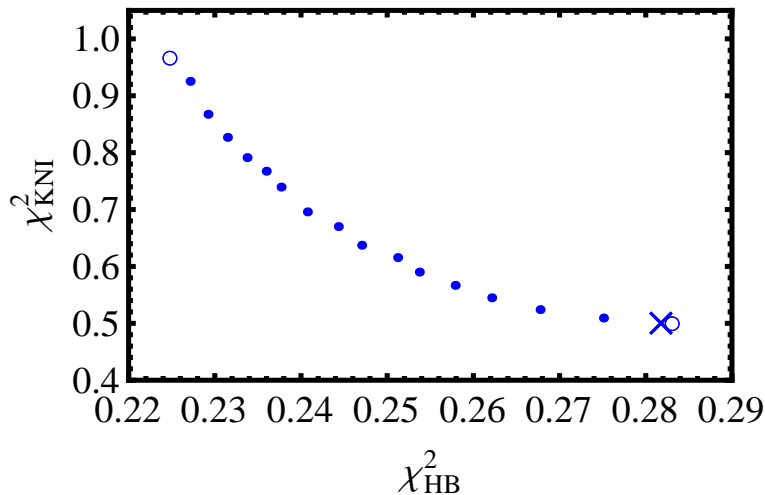}
\caption{Pareto front for the fit of  HB and KNI proteins of Fig.~\ref{fig5}. 
In this bi-objective optimization problem, the coordinates of the fitness space are the reduced chi-squared functions $\chi_{HB}^2(\vec{p}) $ and $\chi_{KNI}^2(\vec{p}) $, where the vector of the parameters $\vec{p}$  is a parameterization of the Pareto front. These functions have been calculated as in (\ref{chi}).
The cross represents the particular instance of the parameter values of Fig.~\ref{fig5}. The circles represent the two other instances of the HB and KNI fits that are shown in Fig.~\ref{fig9}.}
\label{fig8}
\end{figure}

\begin{figure}[htpb]
\centering{
\includegraphics[width=6.5cm]{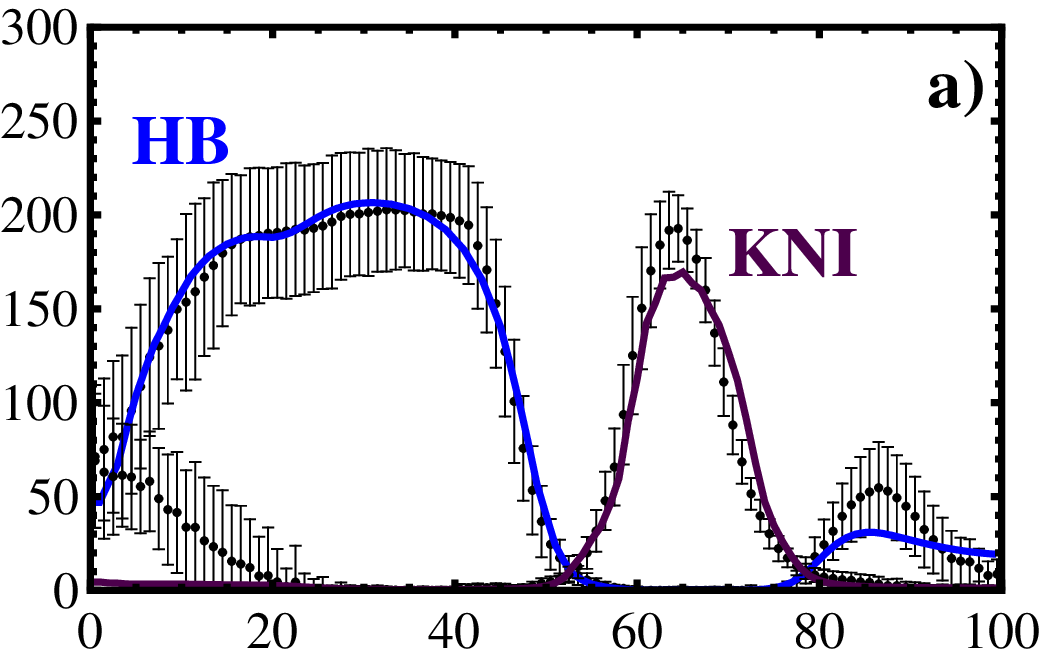} \includegraphics[width=6.5cm]{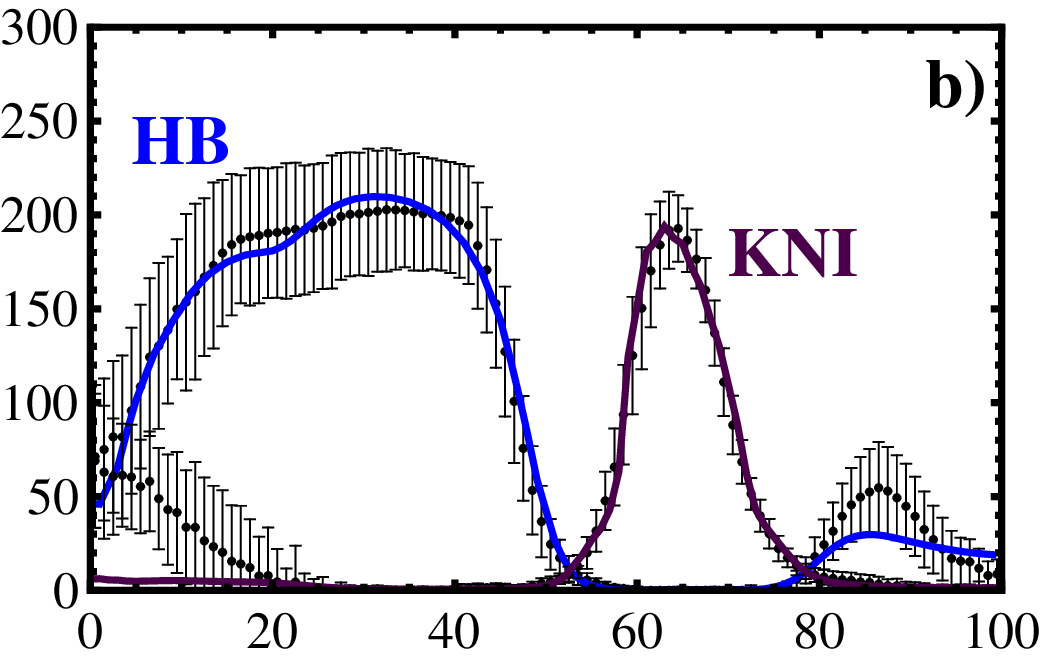}}
\caption{Two instances  of the fit of HB and KNI in the Pareto front, represented by circles in  Fig.~\ref{fig8}. In a), we have the best fit  for HB and the worst fit for KNI. IN b) we have the worst fit HB and the best fit for KNI.}
\label{fig9}
\end{figure}

In  Fig.~\ref{fig8}, we show the Pareto front for the bi-objective optimization problem  associated with the parameter identification describing the distribution of HB and KNI as shown in Fig.~\ref{fig5}. We show the position of the fit of Fig.~\ref{fig5} in the Pareto front of Fig.~\ref{fig8}. In Fig.~\ref{fig9}, we show two other instances of the fits of HB and KNI proteins on the Pareto front. Comparing the three fits, we conclude that they are all acceptable.  

\begin{figure}[htpb]
\centering
\includegraphics[width=10cm]{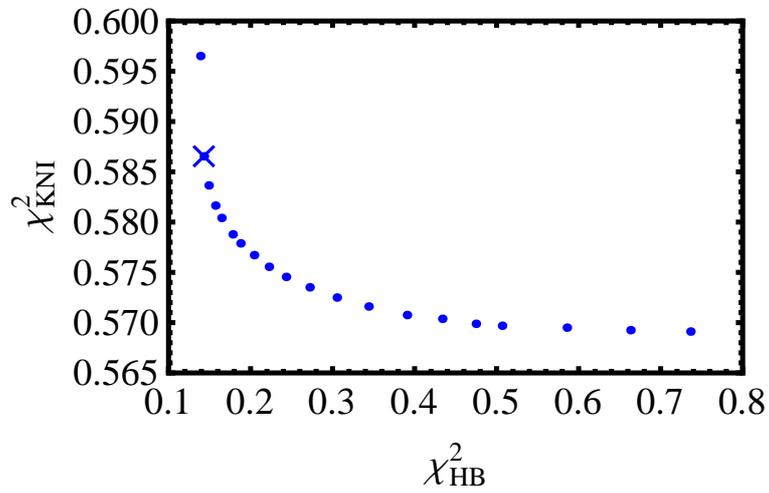}
\caption{Pareto front for the fit of  Hunchback, Knirps and Huckebein proteins of Fig.~\ref{fig7}. The cross represents the particular instance of the parameter values of Fig.~\ref{fig7}.}
\label{fig10}
\end{figure}

In Fig.~\ref{fig10}, we show the Pareto front for the fit of Fig.~\ref{fig7} and we mark the particular instance of the parameters of  Fig.~\ref{fig7}. In all the cases shown here, we conclude that the experimental data are optimally realized by an infinite set of parameters. This is particularly important in biology in the case of  selection pressure affecting simultaneously several phenotypic characteristics of organisms.

\section{Conclusions and Final Remarks}\label{conclusions}

In order to analyze the expression of  the gap gene protein Hunchback  along the antero-posterior axis of the embryo of \textit{Drosophila}, we have introduced a genetic regulatory network model for the proteins HB and KNI and we have calibrated the experimental data with the model predictions. In the most complete version of the model  of Fig.~\ref{fig6}, we have shown that the distribution of HB and KNI along the antero-posterior axis of the embryo are in fact well described by a cross regulation mechanism together with the  transcriptional repression of HKB over HB. We have predicted the distribution of HKB in the form of a localized stripe near the posterior tip of the embryo. Another important conclusion we have obtained is that these patterns are obtained as transient solutions of an ordinary differential equation model, with diffusion playing no role at the level of gap gene protein expression patterns. With this approach, diffusion is only relevant for the establishment of gradients for proteins produced from mRNA with maternal origin. The patterning obtained along the embryo results from the differences in concentrations of the maternal proteins of the embryo.   

The calibration and validation of the genetic regulatory network models have been done with genetic algorithm techniques for parameter identification. We have used single-objective and multi-objective techniques within the genetic algorithms formalism, and we have analyzed the usefulness  of the concept of Pareto optimization in biology. Due to similarities between the fits and the experimental data,  it is plausible to think that, in the presence of several objectives, the number of possible parametric solutions of a given problem is not unique, producing an infinite set of parameter instantiations.   In this framework, the Pareto set and the Pareto front are the correct approach to analyze these problems. In the case of  selection pressure on organisms affecting simultaneously several phenotypic characteristics, the Pareto type solutions appear as the right quantitative approach to quantify phenotypic variability.

In the most difficult case of the multi-objective optimization problem analyzed here,  we have fitted 31 parameters in a system of ordinary differential equations with 18 independent variables, and we have implemented these algorithms in a grid computing environment. In the supplementary material of this paper, we list all the algorithms and all the associated C files developed under this framework, \cite{SUP}.  These techniques are general and can be used in other parameter identification problems.

\section*{Acknowledgments}  This work has been supported  by European project GENNETEC, FP6 STREP IST 034952. The parallel computations in this paper have been done in the Laboratoire de Recherche en Informatic of the INRIA, Universit\'e Paris-Sud, Paris.

\end{document}